\documentclass[conference,twocolumn,10pt]{IEEEtran} 

\usepackage{graphicx}
\usepackage{epstopdf}
\usepackage{amsmath}

\usepackage[linesnumbered,ruled]{algorithm2e}
\usepackage{cite}

\newcommand{\comment}[1]{{}} 

\usepackage{amssymb}
\usepackage{xspace}
\usepackage{bm}

\newcommand{\parens}[1]{\ensuremath{\left(#1\right)}\xspace}
\newcommand{\brackets}[1]{\ensuremath{\left[#1\right]}\xspace}

\newcommand{\bars}[1]{\ensuremath{\left\vert#1\right\vert}\xspace}

\newcommand{\complex}{\ensuremath{\mathbb{C}}\xspace}

\newcommand{\ev}[1]{\ensuremath{\mathbb{E}\brackets{#1}}\xspace}

\newcommand{\distcgauss}[2]{\ensuremath{\mathcal{N}_{\complex}\parens{#1,#2}}\xspace} 

\newcommand{\msnr}{\ensuremath{\mathrm{SNR}}\xspace}

\newcommand{\msinr}{\ensuremath{\mathrm{SINR}}\xspace}

\newcommand{\msir}{\ensuremath{\mathrm{SIR}}\xspace}
\newcommand{\minr}{\ensuremath{\mathrm{INR}}\xspace}

\newcommand{\noisestd}{\ensuremath{\sigma_{\mathrm{n}}}\xspace}
\newcommand{\noisevar}{\ensuremath{\noisestd^2}\xspace}

\newcommand{\Ptx}{\ensuremath{P_{\mathrm{tx}}\xspace}}

\newcommand{\powerdes}{\ensuremath{P_{\mathrm{des}}}\xspace}
\newcommand{\powerint}{\ensuremath{P_{\mathrm{int}}}\xspace}

\newcommand{\bint}{\ensuremath{\bar{b}\xspace}}
\newcommand{\mint}{\ensuremath{\bar{m}\xspace}}
\newcommand{\Omegaint}{\ensuremath{\bar{\Omega}\xspace}}
\usepackage{xcolor}

\usepackage{xspace}
\usepackage[acronym,nogroupskip,nonumberlist,nopostdot]{glossaries}
\makeglossaries

\newacronym{snr}{SNR}{signal-to-noise ratio}
\newacronym{sinr}{SINR}{signal-to-interference-plus-noise ratio}
\newacronym{sir}{SIR}{signal-to-interference ratio}
\newacronym{inr}{INR}{interference-to-noise ratio}
\newacronym{pdf}{PDF}{probability distribution function}
\newacronym{cdf}{CDF}{cumulative distribution function}
\newacronym{leo}{LEO}{low-earth orbit}
\newacronym{frf}{FRF}{frequency reuse factor}
\newacronym{los}{LOS}{line-of-sight}
\newacronym{nlos}{NLOS}{non-line-of-sight}
\newacronym{mimo}{MIMO}{multiple-input multiple-output}
\newacronym{sr}{SR}{Shadowed Rician}
\newacronym{ssr}{SSR}{Squared Shadowed Rician}
\newacronym{5g}{5G}{fifth generation}

\newcommand{\snr}{\gls{snr}\xspace}
\newcommand{\sinr}{\gls{sinr}\xspace}
\newcommand{\sir}{\gls{sir}\xspace}
\newcommand{\inr}{\gls{inr}\xspace}
\newcommand{\los}{\gls{los}\xspace}
\newcommand{\nlos}{\gls{nlos}\xspace}
\newcommand{\sr}{\gls{sr}\xspace}
\newcommand{\ssr}{\gls{ssr}\xspace}
\newcommand{\leo}{\gls{leo}\xspace}
\newcommand{\pdf}{\gls{pdf}\xspace}
\newcommand{\cdf}{\gls{cdf}\xspace}

\newcommand{\figref}[1]{Fig.~\ref{#1}} 

\newcommand{\tabref}[1]{Table~\ref{#1}}
\newcommand{\thmref}[1]{Theorem~\ref{#1}}

\newcommand{\corref}[1]{Corollary~\ref{#1}}

\usepackage{xspace}

\newcommand{\numbeams}{\ensuremath{N_{\mathrm{B}}}\xspace}

\def\b0{{\mathbf{0}}}

\usepackage[caption=false,font=footnotesize]{subfig}

\usepackage{amsthm}

\theoremstyle{plain}
\newtheorem{theorem}{Theorem}

\theoremstyle{plain}
\newtheorem{corollary}{Corollary}[theorem]

\theoremstyle{plain}

\theoremstyle{definition}

\theoremstyle{remark}

\newcounter{mytempeqncnt}

\begin{document}

%
\title{Downlink Analysis of LEO Multi-Beam Satellite Communication in Shadowed Rician Channels}
%
%
%

\author{
    Eunsun Kim, Ian P. Roberts, Peter A. Iannucci, and Jeffrey G. Andrews\\%
    The University of Texas at Austin\\%
    esunkim@utexas.edu%
}


\maketitle

\begin{abstract}
    The coming extension of cellular technology to base-stations in \leo
    requires a fresh look at terrestrial 3GPP channel models.  Relative to such
    models, sky-to-ground cellular channels will exhibit less diffraction,
    deeper shadowing, larger Doppler shifts, and possibly far stronger
    cross-cell interference: consequences of high elevation angles and extreme
    ``sectorization'' of \leo satellite transmissions into partially-overlapping
    spot beams.  To permit forecasting of expected \snr, \inr and probability of
    outage, we characterize the powers of desired and interference signals as
    received by ground users from such a \leo satellite.  In particular,
    building on the \sr channel model, we observe that co-cell and cross-cell
    sky-to-ground signals travel along similar paths, whereas terrestrial co-
    and cross-cell signals travel along very different paths.  We characterize
    \snr, \sir, and \inr using transmit beam profiles and linear relationships
    that we establish between certain \sr random variables.  These tools allow
    us to simplify certain density functions and moments, facilitating future
    analysis.  Numerical results yield insight into the key question of whether
    emerging \leo systems should be viewed as interference- or noise-limited.

\end{abstract}

%
\IEEEpeerreviewmaketitle

\glsresetall 

\section{Introduction} \label{sec:introduction}

LEO satellite communication systems are experiencing a renaissance.  Deployment
costs have dropped dramatically due to new launch technology, both enabling and
being enabled by ongoing mass deployments by companies such as SpaceX and
Amazon.  This renewal of commercial investment in LEO may mark an end to the
commercial-space Dark Ages brought about by the bankruptcies of Iridium and
Globalstar.  Beyond broadband Internet service, LEO communications satellites
are under consideration as components of future cellular systems.

In such a system, each LEO satellite delivers service to ground users through
multiple onboard antennas, collectively producing a number of spot beams that
make up the satellite's coverage footprint.  Due to aggressive frequency reuse
across spot beams, these systems are expected to have nontrivial interference,
making \gls{sir} and \gls{sinr} important metrics rather than simply \gls{snr}
\cite{jointprecoding}\cite{geo_ant}.  Most existing work studying these metrics
to gauge system performance often rely on simple channel models.  For instance,
in \cite{LutzErich2016TtTs}, \snr, \sir, and \sinr of geostationary
satellite-based communication systems are studied
under channels that are deterministic and common across spot beams.

Extensive studies to characterize received signal power on the ground in satellite communication systems have recognized that satellite signals have \los and \nlos components and that both have random fluctuations. 
Among the efforts to parameterize a distribution to fit collected measurements,
the \sr model \cite{newsimple} is a widely accepted distribution that captures the characteristics of downlink signals \cite{sr_simul}. 
Based on the \gls{pdf} of the \sr model, the distribution of a sum of the \gls{ssr} random variables was analyzed in \cite{sumsquaredshadowrician}\cite{closed_sum_ssr} -- assuming the \ssr random variables are independent and uncorrelated -- and then used to model the interference power in satellite communication systems. 
The \gls{cdf} of the \sr distribution is obtained in \cite{closed_sr} in closed-form, however, most of the analytical results in \cite{closed_sr, sumsquaredshadowrician, closed_sum_ssr} involve infinite power series and special functions, which are computationally complex and challenging to gain insight from. 


In this work, we conduct analysis of a LEO satellite communication system with multiple spot beams under the \sr channel model.
Unlike \cite{sumsquaredshadowrician}\cite{closed_sum_ssr}, we assume the downlink desired and interference signal powers are fully correlated rather than independent.  
We justify this by the fact that signals from multiple onboard transmitters reach a ground user after experiencing approximately the same propagation channel due to the vast separation between the satellite and user.
From this, we characterize the desired and interference power levels and important metrics such as \snr, \sir, and \inr.
To do so, we establish a relation between parameters of linearly related \sr random variables.
In addition, we present a useful, yet minor, assumption on \sr fading order that simplifies the density function and moments of \sr random variables.
We provide numerical results to evaluate the system performance of \leo-based satellite communication for various degrees of shadowing and satellite elevation angles, which show that such systems are not necessarily interference-limited, as is the case when only considering simplified channel models that neglect shadowing.

\begin{figure*}[]
    \normalsize
    \setcounter{mytempeqncnt}{\value{equation}}
    \setcounter{equation}{1}
    \begin{gather}
    f_{|h|}(x; b,m,\Omega) 
    = \frac{x}{b} \left(\frac{2b m}{2bm + \Omega}\right)^{m}  \exp\left(-\frac{x^2}{2b}\right) {}_1\mathcal{F}_1 \left(m, 1, \frac{\Omega x^2 }{2b (2bm + \Omega)}\right)
    \label{eq:pdfsr} \\
 	f_{|h|^2}(y; b, m, \Omega) = \frac{1}{2b} \parens{\frac{2bm}{2bm + \Omega}}^{m} \exp\parens{-\frac{ y}{2b}}{}_1\mathcal{F}_1 \parens{m, 1, \frac{\Omega y}{2b (2b m+\Omega)}} \label{eq:ssrpdf} 
    \end{gather}
    \setcounter{equation}{\value{mytempeqncnt}}
    \hrulefill
    \vspace*{4pt}
\end{figure*}

\section{System Model} \label{sec:system-model}

\begin{figure}
	\centering
	\includegraphics[width=0.55\linewidth]{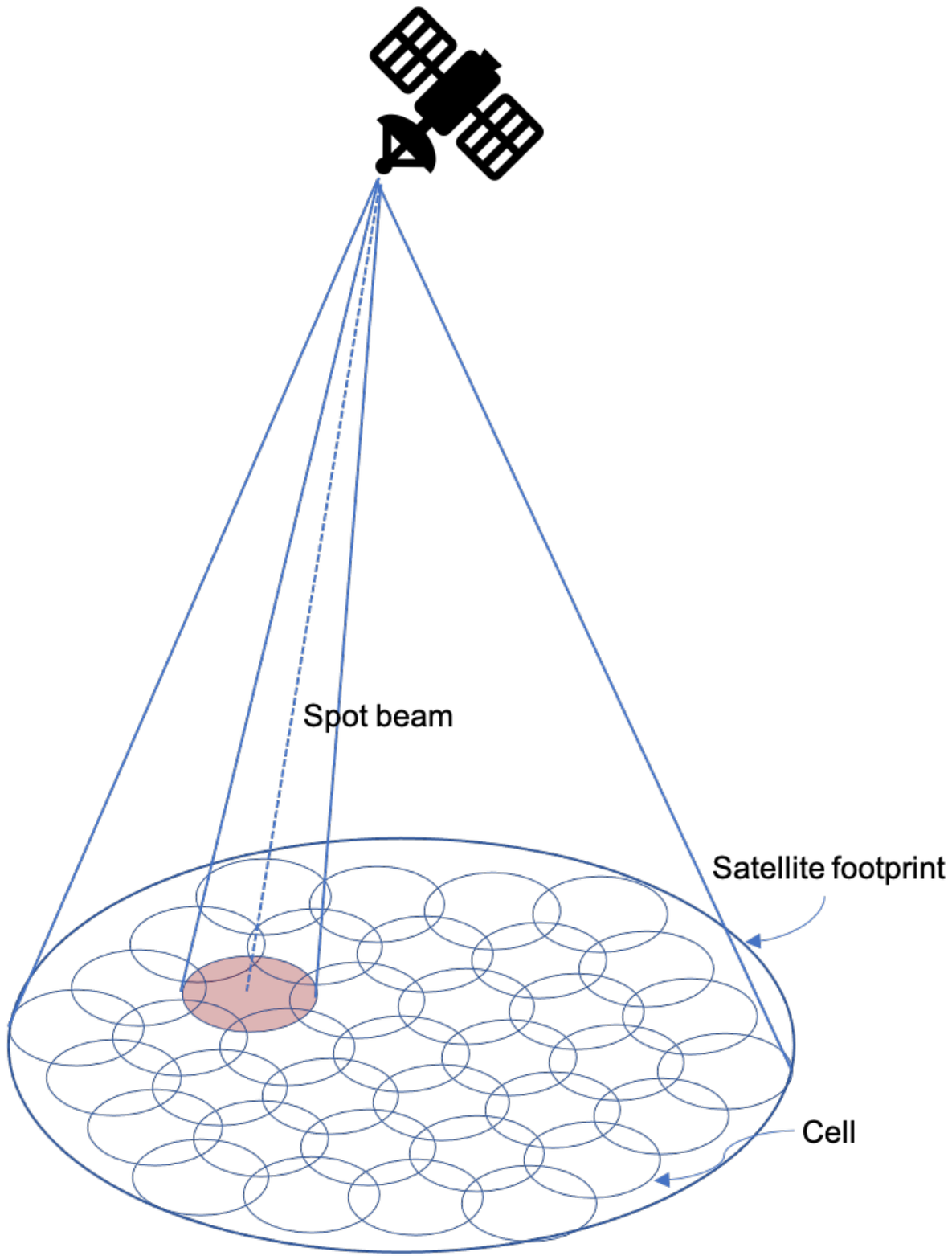}
	\caption{An illustration of the spot beams and their cells comprising the footprint of a satellite communication system.}
	\label{fig:footprint}
\end{figure}
We consider the downlink of ia \gls{leo} satellite communication system serving several ground users simultaneously. 
Multiple transmit antennas at the satellite form \textit{spot beams}, where the area on the ground that each spot beam serves is referred to as a \textit{cell}, as illustrated in \figref{fig:footprint}. 
The coverage area served by all spot beams populate the \textit{satellite footprint}.

Let $\numbeams$ be the number of spot beams the satellite maintains, where each spot beam is created by a dedicated transmitter onboard the satellite, each transmitting with power $\Ptx$.
We assume a frequency reuse factor of one, meaning each spot beam operates using the same time-frequency resource.
As such, $\numbeams - 1$ spot beams inflict interference onto a desired ground user.
Let $x_i$ be the transmitted symbol from the $i$-th onboard transmitter, where $i = 0, \dots, \numbeams-1$.
Let us assume each onboard transmitter is equipped with a dish antenna having gain pattern $G(\theta,\phi)$, where $\theta$ and $\phi$ are the azimuth and elevation relative to boresight, respectively.
We assume the $i$-th transmitting dish is steered to serve the $i$-th cell on the ground.

Given the overwhelming distance between the satellite and a ground user relative to the separation between onboard dish antennas, the desired signal and $\numbeams - 1$ interference signals experience approximately the same propagation channel $h$.
As such, we can write the received symbol of a user being served by the $0$-th spot beam as
\begin{align}
y_0 
=& \ \sqrt{\Ptx \cdot G(\theta_0,\phi_0)} \cdot h \cdot x_0 \nonumber \\
&+ \sum_{i=1}^{\numbeams-1} \sqrt{\Ptx \cdot G(\theta_i,\phi_i)} \cdot h \cdot x_i + n_0,
\label{eq:4}
\end{align}
where $n_0 \sim \distcgauss{0}{\noisevar}$ is additive noise,
$\theta_i$ and $\phi_i$ are elevation and azimuth of a user \textit{relative to boresight} of the $i$-th dish.

\addtocounter{equation}{1} 
\addtocounter{equation}{1} 

We consider a \sr channel model \cite{newsimple}, where $|h| \sim \mathrm{SR}(b,m,\Omega)$ is a \sr random variable with \gls{pdf} 
as in \eqref{eq:pdfsr} where ${}_1\mathcal{F}_1(\cdot, \cdot, \cdot)$ is the confluent hypergeometric function \cite{tablesofintegrals}.
The \sr channel model captures \los and \nlos propagation, like conventional Rician fading, and also incorporates random shadowing into the \los term to account for fluctuations due to the environment.
Intuitively, the model is summarized as $2b$ being the average power of the scattered component, $\Omega$ being the average power of the \los component, and $m$ being the fading order.

\section{Characterizing Downlink Desired and Interference Signal Powers} \label{sec:characterizing-interference}

With our system model in hand, we now characterize desired signal power and the interference inflicted on a ground user by the $\numbeams - 1$ spot beams.
From this, we can gain insights on the distribution of important metrics such as \snr, \sir, \inr and \sinr, which dictate system performance and drive design decisions.

\begin{figure*}[t]
		\setcounter{mytempeqncnt}{\value{equation}}
	
	\setcounter{equation}{8}
	\begin{align}
		f_Y(y)=& f_{|h| = X} \parens{x=\sqrt{\frac{y}{k}}; b,  m, \Omega} \frac{dx}{dy} \label{eq:thm_1-pre} \\
		=& \frac{1}{2kb} \parens{\frac{2kbm}{2kbm + k\Omega}}^{m} \exp\parens{-\frac{ y}{2kb}}{}_1\mathcal{F}_1 \parens{m, 1, \frac{k\Omega y}{2kb (2kb m+k\Omega)}}   \\
		=& f_{|h|^2=Y}(y; kb, m, k\Omega) 
		\label{eq:thm_1}
	\end{align}

	\setcounter{equation}{\value{mytempeqncnt}}
	\hrulefill
\end{figure*}

\subsection{Desired Signal Power}
Henceforth, let us consider a user being served by the $0$-th spot beam.
The \textit{desired} signal power $\powerdes$ received by the user is simply
\begin{align}
\powerdes = \Ptx \cdot G(\theta_0,\phi_0) \cdot |h|^2.
\end{align}
With $|h| \sim \mathrm{SR}(b,m,\Omega)$ modeled as an \sr random variable, $|h|^2$ follows an \ssr (squared \sr) distribution with \gls{pdf} shown in \eqref{eq:ssrpdf} \cite{newsimple}.
Since the received desired signal power is proportional to the channel gain, $\powerdes$ is also an \ssr random variable with distribution related to that of $|h|^2$.
From \thmref{thm:scaled-ssr} below, we conclude that, when $|h|^2 \sim \mathrm{SSR}(b,m,\Omega)$, the received desired signal power is distributed as $\powerdes \sim \mathrm{SSR}(\tilde{b},\tilde{m},\tilde{\Omega})$ where
\begin{align}
    \tilde{b} &= \Ptx \cdot G(\theta_0,\phi_0) \cdot b \\
    \tilde{m} &= m \\
    \tilde{\Omega} &= \Ptx \cdot G(\theta_0,\phi_0) \cdot \Omega
\end{align}

\begin{theorem} \label{thm:scaled-ssr}
    If $X \sim \mathrm{SR}(b,m,\Omega)$ and $Y = k \cdot X^2$, then 
    \begin{align}
    Y \sim \mathrm{SSR}(k \cdot b,m,k \cdot \Omega)
    \end{align}
    \begin{proof}
        See \eqref{eq:thm_1-pre}--\eqref{eq:thm_1} where the \pdf of $Y$ is obtained using the relation $f_Y(y) = f_X(x) \frac{dx}{dy}$ where $f_{|h|}(\cdot)$ and $f_{|h|^2}(\cdot)$ are given in \eqref{eq:pdfsr} and \eqref{eq:ssrpdf}.
    \end{proof}
\end{theorem}

\addtocounter{equation}{3}

\begin{corollary}
    Two \ssr random variables $X \sim \mathrm{SSR}(b_1,m_1,\Omega_1)$ and $Y \sim \mathrm{SSR}(b_2,m_2,\Omega_2)$ are linearly related as $X = k Y$ for $k > 0$ if
    \begin{gather}
    m_1 = m_2 \\
    \frac{b_1}{b_2} = \frac{\Omega_1}{\Omega_2} = k
    \end{gather}
\end{corollary}

The instantaneous \snr observed by the ground user is simply
\begin{align}
\msnr = \frac{\powerdes}{\noisevar} = \frac{\Ptx \cdot G(\theta_0,\phi_0) \cdot |h|^2}{\noisevar}
\end{align}
which, from \thmref{thm:scaled-ssr}, follows an \ssr distribution as
\begin{align} \label{eq:snr-ssr}
    \msnr \sim \mathrm{SSR}\parens{\frac{\Ptx \cdot G(\theta_0,\phi_0)}{\noisevar} \cdot b, m, \frac{\Ptx \cdot G(\theta_0,\phi_0)}{\noisevar} \cdot \Omega}
\end{align}
This may be particularly useful for conducting analyses and simulations involving the probability of outage, ergodic capacity, and the like, especially in settings where interference is low such as with wider cells or less aggressive frequency reuse.



\subsection{Interference Power}
Now, let us turn our attention to characterizing interference.
The total interference power $\powerint$ incurred at the user can be written as
\begin{align}
\powerint 
&= \sum_{i=1}^{\numbeams-1} \Ptx \cdot G(\theta_i,\phi_i) \cdot |h|^2,
\end{align}
which is also merely a scaled version of $|h|^2$ and, thus, of $\powerdes$.
As such, using \thmref{thm:scaled-ssr}, we can readily conclude that \powerint is an \ssr random variable with parameters $\parens{\bint,\mint,\Omegaint}$ as
\begin{align}
    \bint &= \Ptx \cdot \sum_{i=1}^{\numbeams-1} G(\theta_i,\phi_i) \cdot b \\
    \mint &= m \\
    \Omegaint &= \Ptx \cdot \sum_{i=1}^{\numbeams-1} G(\theta_i,\phi_i) \cdot \Omega
\end{align}
While both interference and desired signal power follow the \ssr distribution, it is important to realize that they are fully correlated since both are scaled versions of the channel gain $|h|^2$; recall this is due to the fact that the stochastics seen by the signal of the $0$-th spot beam are also seen by the signals of the other spot beams considering they propagate along the same path to the user.
With this being said, the \sir can be expressed \textit{deterministically} as
\begin{align}
\msir = \frac{\powerdes}{\powerint} = \frac{G(\theta_0,\phi_0)}{\sum_{i=1}^{\numbeams-1} G(\theta_i,\phi_i)}
\end{align}
In other words, we see that \sir is completely irrespective of the shadowing realization and is solely a function of the position of a user relative to the steering direction of the onboard dish antennas.
The \sinr observed by a user, on the other hand, is indeed a random variable tied to the shadowing distribution, though it is challenging to characterize statistically. 
A useful expression of \sinr as a function of \gls{snr} and \gls{inr} is
\begin{align}
	\msinr = \frac{\powerdes}{\noisevar + \powerint} = \frac{\msnr}{1+\minr} 
\end{align}
where $\minr = {\powerint}/{\noisevar}$ is also an \ssr random variable.
The received \inr is a convenient metric gauging if a system is \textit{noise}-limited or \textit{interference}-limited, which can drive cell placement/size and other system factors such as antenna beam pattern.
Realizing that $\minr$ is an \ssr random variable whose distribution is tied to system parameters can provide engineers with means to better tailor satellite communication systems in the face of shadowing.


\begin{figure*}[t]
    \normalsize
    \setcounter{mytempeqncnt}{\value{equation}}
    \setcounter{equation}{21}
    \begin{align}
    \tilde{f}_Y(y;b,m,\Omega) =&\frac{1}{2b} \left(\frac{2b m}{2b m + \Omega}\right)^{m}   \exp\parens{-\frac{my}{ 2b m + \Omega }}  
    \sum_{i=0}^{m -1} \frac{(m - 1)! }{(m-1-i)!(i!)^2}\parens{\frac{\Omega y}{2b (2b m + \Omega) }}^i
    \label{eq:integer-pdf} \\
    \tilde{F}_Y(y) 
    =& \frac{1}{2b} \left(\frac{2b m}{2b m + \Omega}\right)^{m} \int_0^y \exp\parens{-\frac{m}{ 2b m + \Omega }t} 
    \sum_{i=0}^{m -1} \frac{(m - 1)! }{(m-1-i)!(i!)^2}\parens{\frac{\Omega t }{2b (2b m + \Omega) }}^i dt \label{eq:integer-cdf-begin} \\
    =&\frac{1}{2b} \left(\frac{2b m}{2b m + \Omega}\right)^{m}  
    \sum_{i=0}^{m -1} \frac{(m - 1)! }{(m-1-i)!(i!)^2}\parens{\frac{\Omega }{2b (2b m + \Omega) }}^i 
    \int_0^y \exp\parens{-\frac{m}{ 2b m + \Omega }t}  t^i dt \\
    =&\left(\frac{2b m}{2b m + \Omega}\right)^{m-1}  
    \sum_{i=0}^{m -1} \frac{(m - 1)! }{(m-1-i)!(i!)^2}\parens{\frac{\Omega }{2bm}}^i  \parens{\Gamma(i+1) - \gamma\parens{i+1, \frac{m y}{2 b m + \Omega}}} 
    \label{eq:integer-cdf}
    \end{align}
    \setcounter{equation}{\value{mytempeqncnt}}
    \hrulefill
    \vspace*{4pt}
\end{figure*}



\begin{table}[!t]
    \small
    \centering
    \caption{Desired received signal power \ssr parameters used in \cite{newsimple}\cite{sr_simul}.}
    \label{table:channel}
    \begin{tabular}{c|c|c|c}   
        \textit{Shadowing} &  $\tilde{b}$ & $\tilde{m}$ & $\tilde{\Omega}$ \\
        \hline
        Light & 0.158 & 19.4 & 1.29 \\
        \hline	
        Average& 0.126 & 10.1 & 0.835 \\
        \hline
        Heavy& 0.063 & 0.739 & 8.97 $\times 10^{-4}$\\
    \end{tabular}	
\end{table}

\section{A Useful Assumption on Fading Order} \label{sec:fading-order}

While it is fairly straightforward to show that desired signal power and interference power follow \ssr distributions, the complex nature of such a distribution---chiefly the infinite series in the Confluent Hypergeometric function ${}_1\mathcal{F}_1(\cdot, \cdot, \cdot)$---makes it complicated to implement numerically and difficult to assess analytically.
This motivates an approximation on the distribution that offers convenience in both numerical realization and mathematical analysis.

Let us further motivating this by the fact that satellite channel measurements (e.g., \cite{newsimple}\cite{loo}) have been collected and fitted to the \ssr distribution and have become ubiquitous in literature as realistic shadowing parameters under various settings;
these papers suggest the shadowing parameters as in \tabref{table:channel} to categorize shadowing conditions as light, average, or heavy, from which engineers can simulate, evaluate, and tailor satellite communication systems.
Note that the shadowing parameters in \tabref{table:channel} are denoted with tildes since they correspond to measurements of \textit{received signal power} rather than to the channel itself.
We would like to point out, however, that these received power shadowing parameters can be mapped to channel parameters by scaling $\tilde{b}$ and $\tilde{\Omega}$ according to \thmref{thm:scaled-ssr} to recover the channel parameters $b$ and $\Omega$.

The \ssr parameters in \tabref{table:channel} are certainly useful but do not provide much intuition (e.g., in terms of expected received signal power) nor are they easily used in analysis or perfectly realized numerically.
To address this, we utilize the following theorem from \cite{mathhandbook}.


\begin{figure}[!t]
    \centering
    \includegraphics[width=\linewidth,height=\textheight,keepaspectratio]{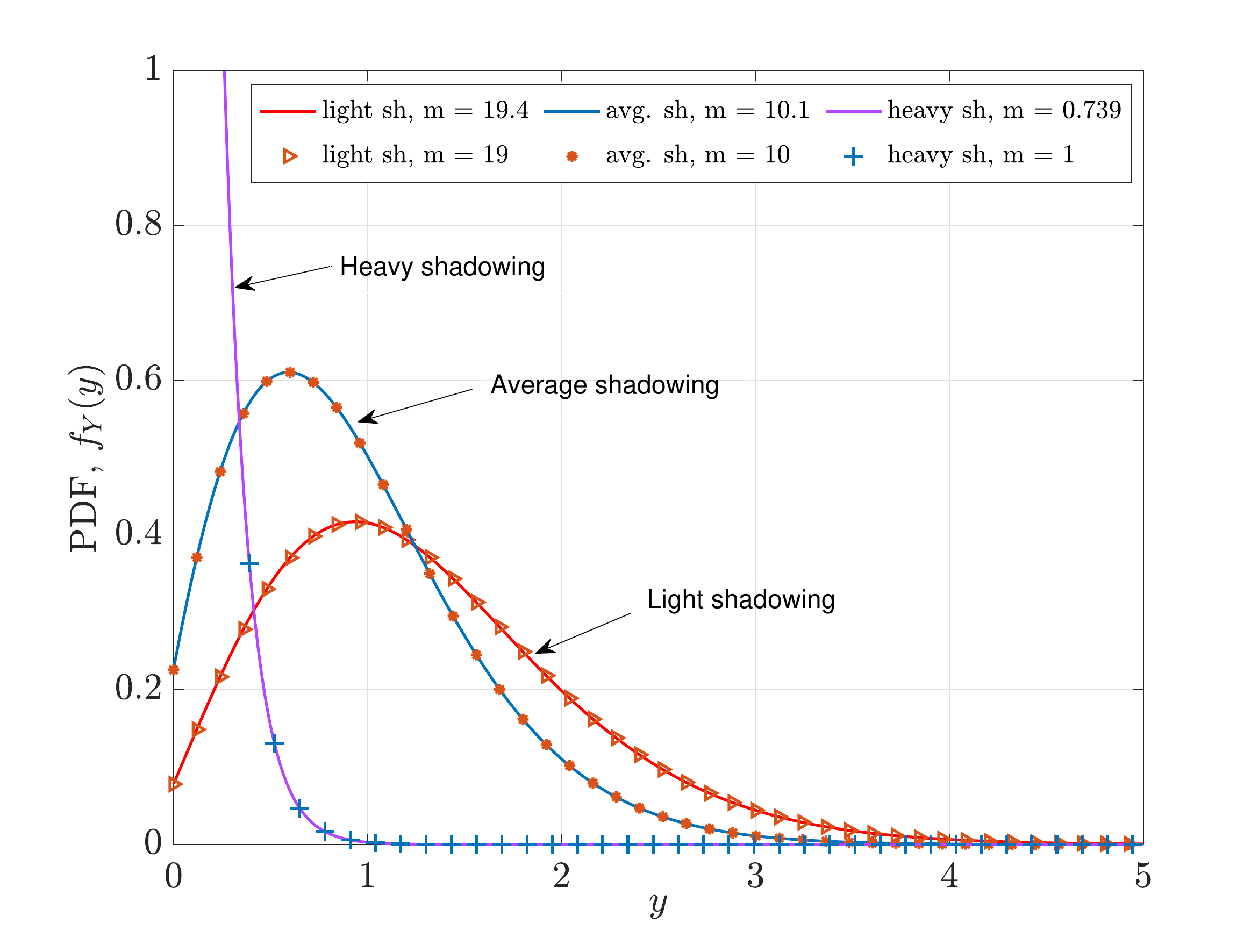} 
    \caption{The \gls{pdf} of an \ssr random variable $Y$ with integer and non-integer fading order $m$ for various levels of fading.}
    \label{fig:integer_m}
\end{figure}


\begin{theorem} \label{thm:m-integer}
    An \ssr random variable $Y \sim \mathrm{SSR}(b,m,\Omega)$ has \gls{pdf} in \eqref{eq:integer-pdf} and \gls{cdf}  in \eqref{eq:integer-cdf-begin}--\eqref{eq:integer-cdf} when $m$ is an integer, 
    where $\Gamma(i+1) = i!$ for integer $i$ and $\gamma(a,x) = \int_{0}^{x} e^{-t}t^{a-1} \ dt$ is the unnormalized incomplete Gamma function \cite{mathhandbook}.
    
    \addtocounter{equation}{4} 
    
    \begin{proof}
        Via Kummer's transform \cite{mathhandbook} as a polynomial, we have
        \begin{align}
        {}_1\mathcal{F}_1(m, 1,\Omega)  
        =e^{\Omega}\sum_{i=0}^{m-1} \frac{(m-1)! \cdot \Omega^i}{(m-1-i)! \cdot (i!)^2}
        \label{eq:confluent_sim}
        \end{align}
        which, along with algebra, yields \eqref{eq:integer-pdf} and \eqref{eq:integer-cdf}.
    \end{proof}
\end{theorem}

\figref{fig:integer_m} shows the \pdf of an \ssr random variable $Y$ with integer and non-integer $m$ for various levels of shadowing---particularly, those in \tabref{table:channel}.
It can be seen that rounding $m$ to an integer in any of the three shadowing levels does not result in noticeable changes in distribution.
Rounding $m$ to an integer will offer convenience analytically, as we will highlight further, and makes little difference in the distribution versus non-integer $m$; this is especially true when \ssr parameters are fitted from measurements since no channel model will \textit{perfectly} represent the true channel.

With \thmref{thm:m-integer}, we can exactly numerically represent the \gls{pdf} and \gls{cdf} of an \ssr random variable and can more easily analyze it mathematically.
For instance, when $m$ is not an integer, the mean of $Y$ is quite involved to express, involving the Gauss Hypergeometric function \cite{tablesofintegrals}. 
However, when $m$ is an integer, the mean becomes quite easy to express and provides straightforward consequences of $b$ and $\Omega$.

\begin{corollary} \label{cor:mean-ssr}
    The mean of $Y \sim \mathrm{SSR}(b,m,\Omega)$ when $m$ is an integer is
    \begin{align}
    \ev{Y} = 2 \cdot b + \Omega 
    \label{eq:corol-mean}
    \end{align}
    \begin{proof}
        \eqref{eq:corol-mean} is obtained using  the formula
        $\int_{0}^{\infty} z^i \cdot e^{-\mu z} \ dz= i! \cdot \mu^{-i-1}$ in \cite{tablesofintegrals}.
    \end{proof}
\end{corollary}

\begin{figure*}[!t]
    \centering
    \subfloat[No shadowing.]{\includegraphics[width=0.5\linewidth]{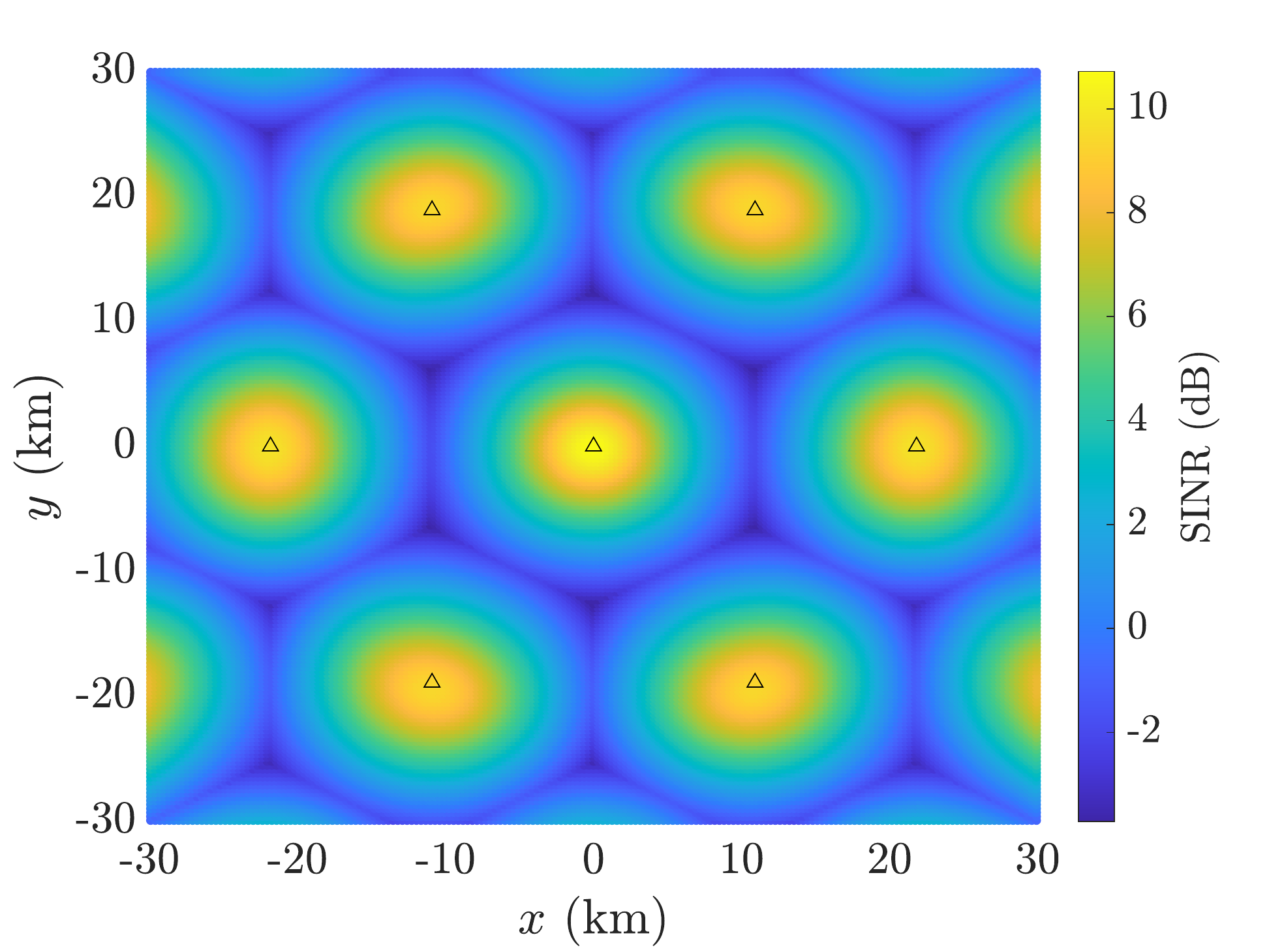} \label{fig:sinr-heatmap-a}}
    \subfloat[Average shadowing.]{\includegraphics[width=0.5\linewidth]{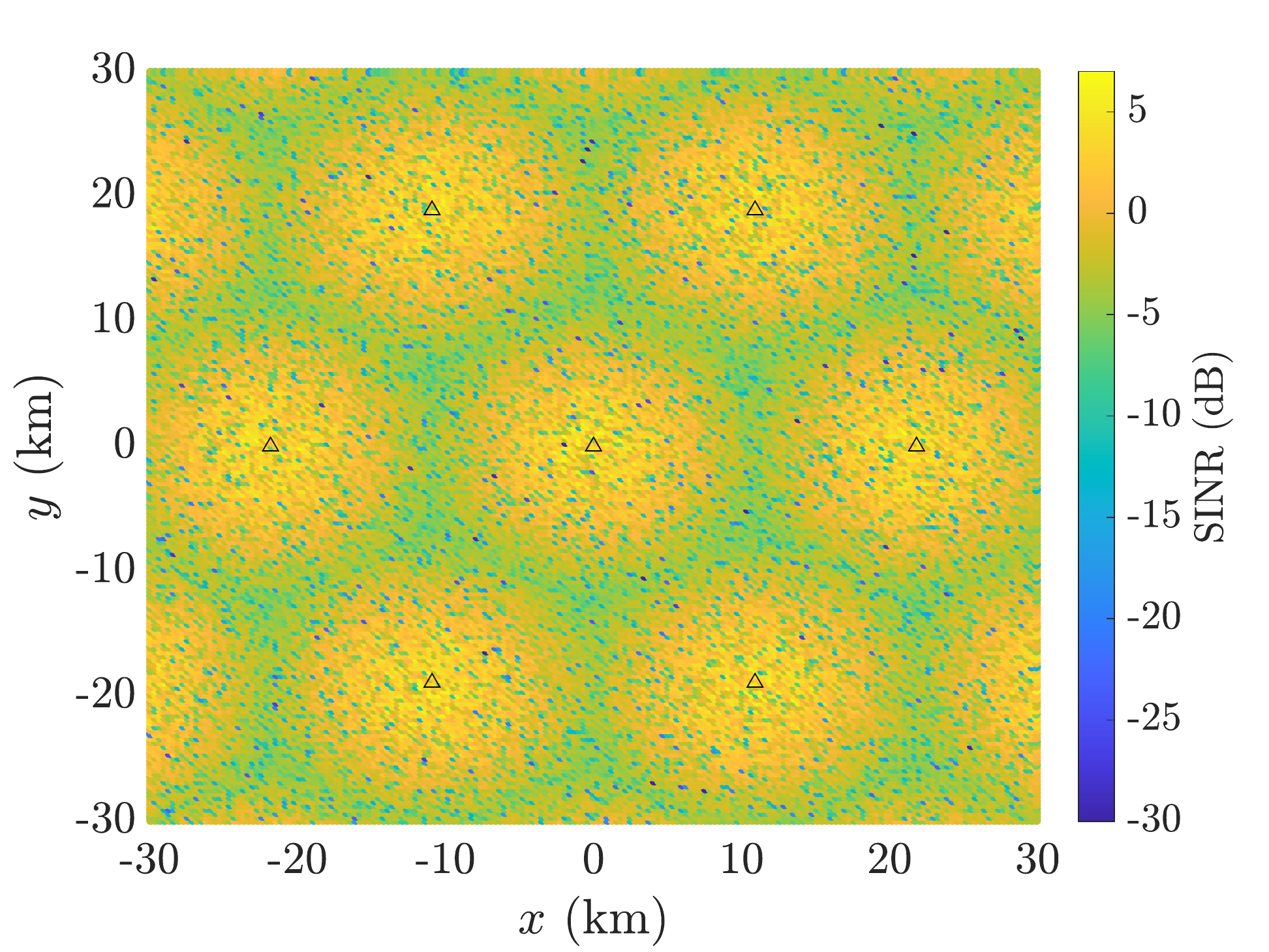} \label{fig:sinr-heatmap-b}}
    \caption{A realization of \sinr as a function of ground user location (a) with no shadowing and (b) with average shadowing, where the satellite elevation angle is $90^\circ$. Triangles indicate the location of cell centers.}
    \label{fig:sinr-heatmap}
\end{figure*}

The average \snr with integer $m$,  using \corref{cor:mean-ssr} and \eqref{eq:snr-ssr} is
\begin{align}
\ev{\msnr} = \frac{\Ptx \cdot G(\theta_0,\phi_0)}{\noisevar} \cdot \parens{2 \cdot b + \Omega}.
\end{align}
The average of other metrics such as $\minr$, $\powerdes$, and $\powerint$, which are all \ssr random variables, can be expressed similarly using \corref{cor:mean-ssr}.

In addition to expected values, the \cdf of an \ssr random variable with an integer $m$ is useful for a variety of communication system performance metrics.  For instance, probability of SNR outage
\begin{align}
	\mathbb{P}\parens{\msnr \leq \gamma} = \tilde{F}_Y\parens{\gamma \cdot \frac{\noisevar}{\Ptx \cdot G(\theta_0,\phi_0)} }
\end{align}
can be evaluated using \thmref{thm:m-integer}.

\section{Numerical Results} \label{sec:simulation-results}

We simulate a \leo-based satellite communication system to highlight the importance and consequences of \sr channels on desired signal and interference powers as well as on \snr, \sir, \inr, and \sinr.
We simulate a satellite communication system at $500$ km altitude with $\numbeams = 19$ spot beams operating at $2$ GHz over a bandwidth of $30$ MHz at various elevation angles.
The spot beams create $19$ cells on the ground in a hexagonal fashion, each with a cell radius of $12.6$ km (corresponding to the $3$ dB beamwidth at an elevation of $90^\circ$).
Each spot beam is steered toward its corresponding cell center using a $30$ dBi dish antenna whose azimuth and elevation patterns \cite{3gpp38811} are
\begin{align}
g(\zeta) = 
\begin{cases}
1, & \zeta = 0^\circ \\
4 \bars{\frac{J_1(ka \sin \zeta)}{ka \sin \zeta}}^2, & 0^\circ < |\zeta| \leq 90^\circ
\end{cases}
\label{eq:beam}
\end{align}
where $\zeta$ is the azimuth or elevation angle, $J_1(\cdot)$ is the first-order Bessel function of the first-kind, $a$ is the radius of the dish antenna, $k = 2 \pi / \lambda$ is the wave number, and $\lambda$ is the carrier wavelength; we simulate dish antennas having radius $a = 10 \lambda$.
The resulting antenna gain at some $(\theta,\phi)$ can be written as $G(\theta,\phi) = g(\theta) \cdot g(\phi)$.
Each onboard transmitter supplies $4$ dBW/MHz of power. 
We randomly distribute ground users across satellite cells, each of which has a noise power spectral density of $-167$ dBm/Hz and a receive antenna gain of $0$ dBi.
When simulating shadowing, we employ the light, average, and heavy shadowing conditions in \tabref{table:channel}. 
In addition to free space path loss, $5.3$ dB of path loss including scintillation loss, atmospheric loss, and shadowing margin are considered. 



We begin by considering the result of \figref{fig:sinr-heatmap}, where we have plotted the realized \sinr as a function of ground user location for the cases of no shadowing in \figref{fig:sinr-heatmap-a} and with average shadowing in \figref{fig:sinr-heatmap-b}.
For the sake of discussion, let us consider only the center cell and the six surrounding cells.
In the case of no shadowing, we can see that the \sinr at a given location is deterministic based on our hexagonal cell layout.
In any of these seven cells, users near the center of the cell see higher \sinr and those near the cell edge see lower \sinr due to interference from neighboring spot beams and the fact that spot beams are steered to cell centers.
In the case of average shadowing, \sinr is no longer well-defined as a function of user location.
Instead, the stochastics of shadowing can play a significant role in the level of desired signal power and interference power that a user receives.
Even users close to the cell center are susceptible to low \sinr, though they \textit{tend} to see higher \sinr than those on the cell edge.
With average shadowing, extremely poor \sinr levels---on the order of $-20$ dB or more---are not unlikely.
These results highlight the potential for poor signal quality even when near the cell center and the very deep fades that are not practical for communication.

\begin{figure}[!t]
    \centering
    \includegraphics[width=\linewidth,height=\textheight,keepaspectratio]{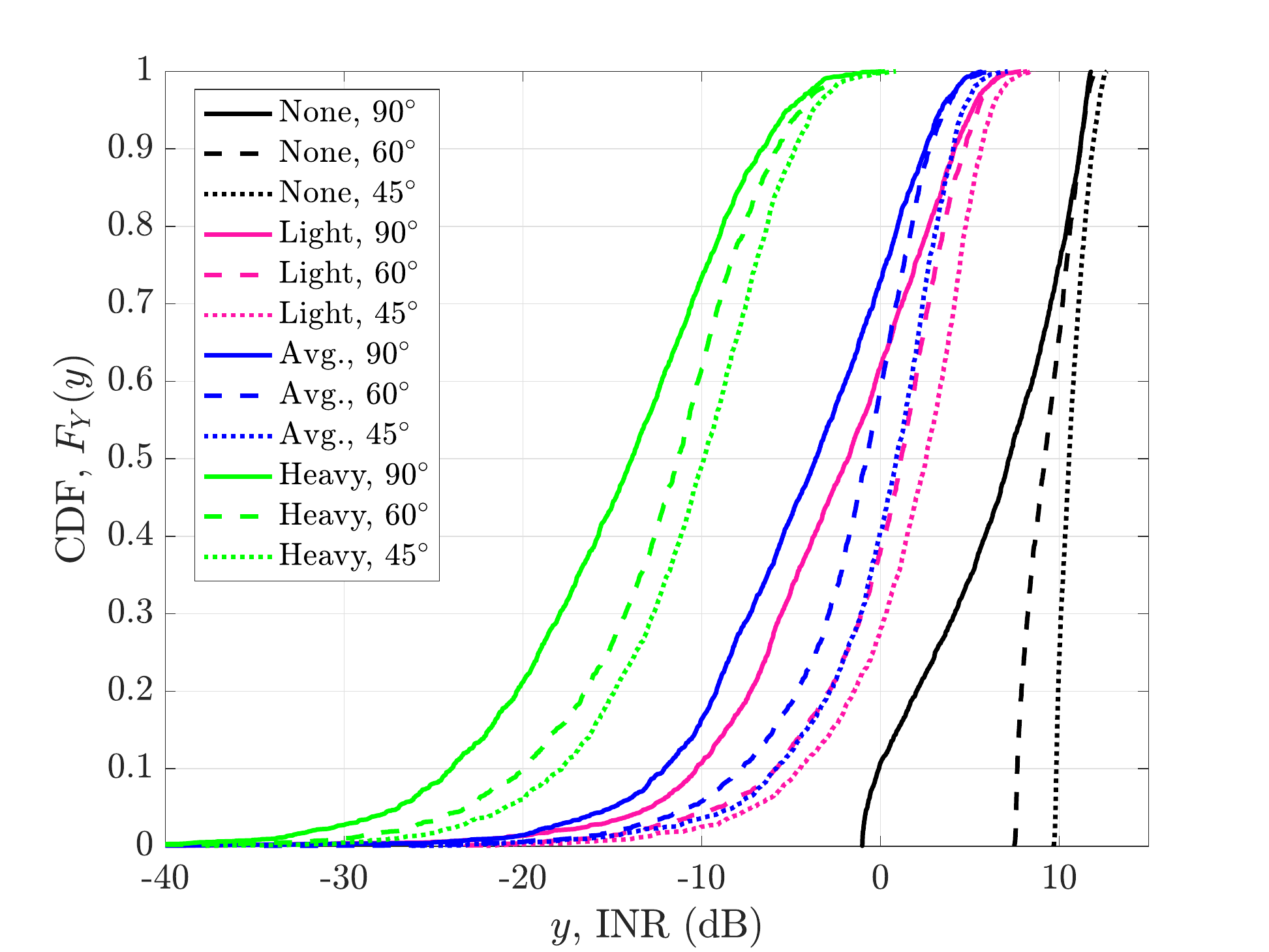}
    \caption{The \glspl{cdf} of \inr across the center cell under various levels of shadowing and elevation angles.}
    \label{fig:cdf-inr-snr-a}
\end{figure}

In \figref{fig:cdf-inr-snr-a}, we look at the \gls{cdf} of \inr, evaluated at user locations across the center cell, for elevation angles of $90^\circ$, $60^\circ$, and $45^\circ$.
Recall that \inr is a useful metric for gauging if a system is \textit{noise}-limited ($\minr \ll 0$ dB) or \textit{interference}-limited ($\minr \gg 0$ dB).
Without shadowing, we can see that almost all users across the cell have a high \inr at all three elevation angles, suggesting that they are interference-limited. 
As a result, \sinr can be well approximated by \sir and noise can be neglected.
With shadowing, however, it is clear that the system is no longer necessarily interference-limited.
Under light and average shadowing, we can see that there are significant probabilities of $\minr$ being above or below $0$ dB.
However, the tails below $0$ dB are much more substantial than those above, which only reaches at most around $8$ dB, suggesting that even then the system is not in an overwhelmingly interference-limited regime.
With heavy shadowing, the \inr is almost exclusively less than $0$ dB, with a significant density below $10$ dB.
These results can be explained by the fact that, in addition to shadowing diminishing the power of the \textit{desired} receive signal, it also reduces that of \textit{interference}.
As the elevation angle decreases from directly overhead at $90^\circ$, the main lobes of the spot beams begin to overlap, driving up inter-cell interference and, thus, the \inr.

\begin{figure}
	\centering
	\includegraphics[width=1\linewidth]{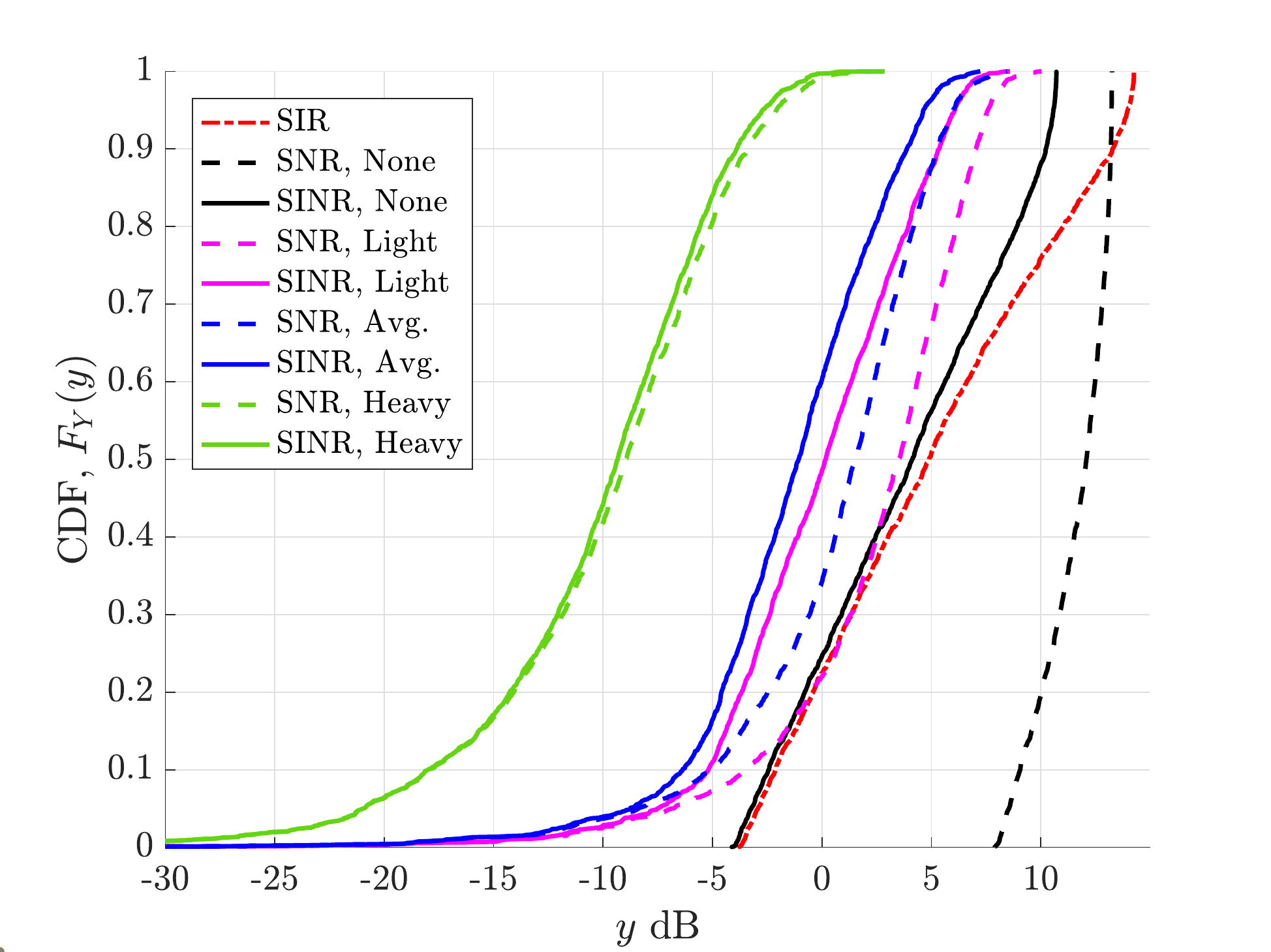} 
	\caption{The \glspl{cdf} of \sir, \snr, and \sinr across the center cell at an elevation of $90^\circ$.}
	\label{fig:cdf-shadowing}
\end{figure}

\figref{fig:cdf-shadowing} shows the \glspl{cdf} of \sir, \snr, and \sinr under various shadowing levels at an elevation of $90^\circ$. 
Recall that \sir is invariant of the shadowing level and, thus, only a single line is shown.
Without shadowing, there is a sizable gap between \snr and \sinr since the system is interference-limited, as discussed.
With light and average shadowing, the gap between \snr and \sinr shrinks, though it remains significant.
With heavy shadowing, however, the gap nearly vanishes as the system becomes fully noise-limited and \sinr is well approximated by \snr.

\section{Conclusion} \label{sec:conclusion}

We have characterized the desired (co-cell) and interference (cross-cell)
signal power levels of a cellular satellite communication system downlink under
a \sr channel model.  Leveraging the fact that co-cell and cross-cell signals
from spot beams travel along the same propagation channel, we have obtained
distributions for received signal power along with \snr, \sir, and \inr, where
we saw that \sir is independent of fading.  We have established means to relate
the parameters of linearly related \sr variables, and we have shown numerically
that approximating the fading order to the nearest integer offers convenience
in analysis at little cost in generality.  The main benefit of integer fading
order lies in the development of tractable algebraic expressions for modeled
probability densities and expected values.  Finally, we have provided numerical
evidence that sky-to-ground cellular channels, though potentially
interference-limited due to limited cross-cell isolation, are in fact rendered
noise-limited when realistic levels of shadowing are introduced.

\bibliographystyle{IEEEtran}
\bibliography{refs}

\begin{thebibliography}{10}
\providecommand{\url}[1]{#1}
\csname url@samestyle\endcsname
\providecommand{\newblock}{\relax}
\providecommand{\bibinfo}[2]{#2}
\providecommand{\BIBentrySTDinterwordspacing}{\spaceskip=0pt\relax}
\providecommand{\BIBentryALTinterwordstretchfactor}{4}
\providecommand{\BIBentryALTinterwordspacing}{\spaceskip=\fontdimen2\font plus
\BIBentryALTinterwordstretchfactor\fontdimen3\font minus
  \fontdimen4\font\relax}
\providecommand{\BIBforeignlanguage}[2]{{%
\expandafter\ifx\csname l@#1\endcsname\relax
\typeout{** WARNING: IEEEtran.bst: No hyphenation pattern has been}%
\typeout{** loaded for the language `#1'. Using the pattern for}%
\typeout{** the default language instead.}%
\else
\language=\csname l@#1\endcsname
\fi
#2}}
\providecommand{\BIBdecl}{\relax}
\BIBdecl

\bibitem{jointprecoding}
B.~{Devillers}, A.~{Perez-Neira}, and C.~{Mosquera}, ``Joint linear precoding
  and beamforming for the forward link of multi-beam broadband satellite
  systems,'' \emph{Proc. {IEEE} Global Commun. Conf.}, Dec. 2011.

\bibitem{geo_ant}
A.~{Kyrgiazos}, B.~{Evans}, P.~{Thompson}, P.~{Mathiopoulos}, P.~{Takis}, and
  S.~{Papaharalabos}, ``A terabit/second satellite system for {European}
  broadband access: A feasibility study,'' \emph{Intl. J. Sat. Commun. Net.},
  vol.~32, pp. 63--92, Mar. 2014.

\bibitem{LutzErich2016TtTs}
E.~{Lutz}, ``\BIBforeignlanguage{eng}{Towards the {Terabit/s} satellite -
  interference issues in the user link},'' \emph{\BIBforeignlanguage{eng}{Intl.
  J. Sat. Commun. Net.}}, vol.~34, pp. 461--482, June 2015.

\bibitem{newsimple}
A.~{Abdi}, W.~C. {Lau}, M.~{Alouini}, and M.~{Kaveh}, ``A new simple model for
  land mobile satellite channels: first- and second-order statistics,''
  \emph{IEEE Trans. Wireless Commun.}, vol.~2, no.~3, pp. 519--528, May 2003.

\bibitem{sr_simul}
M.~R. Bhatnagar and A.~M.K., ``On the closed-form performance analysis of
  maximal ratio combining in {Shadowed-Rician} fading {LMS} channels,''
  \emph{{IEEE} Commun. Lett.}, vol.~18, no.~1, pp. 54--57, Jan. 2014.

\bibitem{sumsquaredshadowrician}
G.~{Alfano} and A.~{De Maio}, ``Sum of squared {Shadowed-Rice} random variables
  and its application to communication systems performance prediction,''
  \emph{IEEE Trans. Wireless Commun.}, vol.~6, no.~10, pp. 3540--3545, Oct.
  2007.

\bibitem{closed_sum_ssr}
M.~C. Clemente and J.~F. Paris, ``Closed-form statistics for sum of squared
  {Rician} shadowed variates and its application,'' \emph{Electronics Letters},
  vol.~50, pp. 120--121, Jan. 2014.

\bibitem{closed_sr}
J.~F. {Paris}, ``Closed-form expressions for {Rician} shadowed cumulative
  distribution function,'' \emph{Electronics Letters}, vol.~46, pp. 952--953,
  Jul. 2010.

\bibitem{tablesofintegrals}
S.~Gradshteyn and I.~M. Ryzhik, \emph{Table of Integrals, Series, and
  Products}.\hskip 1em plus 0.5em minus 0.4em\relax New York: Academic, 2000.

\bibitem{loo}
C.~{Loo}, ``A statistical model for a land mobile satellite link,'' \emph{IEEE
  Trans. Veh. Technol.}, vol.~34, no.~3, pp. 122--127, Aug. 1985.

\bibitem{mathhandbook}
M.~Abramowitz and I.~A. Stegun, \emph{Handbook of Mathematical
  Functions}.\hskip 1em plus 0.5em minus 0.4em\relax New York: Dover
  Publications, 1994.

\bibitem{3gpp38811}
3GPP, ``{Technical Report} 38.811, {Study} on {New Radio (NR)} to support
  non-terrestrial networks {(NTN)},'' Jul. 2020.

\end{thebibliography}

\end{document}